\documentclass[a4paper,11pt]{article}

\usepackage{bm}
\usepackage{booktabs}
\usepackage{array}
\usepackage{latexsym}  
\usepackage{dcolumn}
\usepackage{amsmath,amsfonts,amssymb}
\usepackage{graphicx,epsfig}
\usepackage{color}
\usepackage{verbatim}
\usepackage{psfrag}
\usepackage{cite}
\usepackage{mathrsfs}

\def\app#1{{Appendix~\ref{#1}}}
\def\sec#1{{Section~\ref{#1}}}
\def\eq#1{{Eq.~(\ref{#1})}}

\def\dd{\mathrm{d}}

\title{Quantum field theory in de Sitter and quasi-de Sitter spacetimes: Revisited}

\author{Suprit~Singh$^a$\footnote{suprit@iucaa.ernet.in},~Chandrima Ganguly$^b$\footnote{ganguly.chandrima@gmail.com}~~and T.~Padmanabhan$^a$\footnote{nabhan@iucaa.ernet.in}\\
\\
$^a$IUCAA, Ganeshkhind, Pune 411007, INDIA. \\ 
\\
$^b$Department of Physics, IIT Hyderabad, Yeddumailaram 502 205, INDIA.\\ 
} 

\date{}

\begin{document}
\maketitle 

\begin{abstract}
It is possible to associate temperatures with the non-extremal horizons of a large class of spherically symmetric spacetimes using periodicity in the Euclidean sector and this procedure works for the de Sitter spacetime as well.
But, unlike e.g., the black hole spacetimes, the de Sitter spacetime also allows a description in Friedmann coordinates. This raises the question of whether the thermality of the de Sitter horizon can be obtained, working entirely in the Friedmann coordinates, without reference to the static coordinates or using the symmetries of de Sitter spacetime.
We discuss several aspects of this issue for de Sitter and approximately de Sitter spacetimes,
in the Friedmann coordinates (with a time-dependent background and the associated ambiguities in defining the vacuum states). The different  choices for the vacuum states, behaviour of the mode functions and the detector response are studied in both (1+1) and (1+3) dimensions. We compare and contrast the differences brought about by the different choices. In the last part of the paper, we also describe a general procedure for studying quantum field theory in spacetimes which are \textit{approximately} de Sitter and, as an example, derive the corrections to thermal spectrum due to the presence of pressure-free matter.
\end{abstract}

\section{Introduction} \label{sec:intro}

There exists an extensive literature on quantum field theory in de Sitter spacetimes (for a non exhaustive sample, see Ref.~\cite{lit1,lit2}). From a theoretical point of view, the high level of symmetry exhibited by the de Sitter geometry makes it an important and tractable example. On the other hand, observations suggest that the evolution of our universe is described by a (near) de Sitter geometry both during the early inflationary phase as well as during the current accelerated phase of expansion. While quantum effects are not expected to play a serious role in the current phase  of the expansion (see, however, \cite{tpcc}), they play an important role during the inflationary phase and possibly seeds the cosmic structure we see today. This was part of the motivation to study quantum field theory in de Sitter background.

 A key feature which arises in such a study of the de Sitter spacetime is the thermal nature of the vacuum state and the  temperature which one can associate with the horizon. One of the purposes of this paper is to explore aspects of this thermality from different perspectives. Since it is a fairly well trodden path, we will begin by describing the motivation (for yet another paper!) and the specific point of view adopted here. 

The connection between spacetime horizons and thermodynamics is quite well-known. The first example of this kind, of course, was in the case of black hole spacetimes \cite{beken-hawking}. This was soon followed by the discovery of horizon temperature in much wider contexts like, for example, in the case of Rindler spacetime \cite{davies-unruh} and de Sitter spacetimes \cite{gibbons}. While all these three spacetimes (black hole, Rindler, de Sitter) have very similar metric structure --- when one uses static coordinates which cover part of the manifold --- they also have significant differences \cite{TPdS}. For example, one can provide a fairly rigorous geometrical description of the black hole horizon. In contrast, the horizon in the Rindler spacetime is just a null surface in flat spacetime as viewed by uniformly accelerated observers and is clearly observer dependent. The situation as regard de Sitter spacetime is somewhat in between: While it has a geometric description, it also shares with the Rindler spacetime certain level   of observer dependence.  The location of the horizon surface will be different even for two different observers who are translated with respect to each other by a spatial vector.  More importantly, the de Sitter spacetime allows, in addition to the static coordinate system,  a spatially homogeneous Friedmann coordinate description. In these coordinates, quantum field theory reduces to the quantum mechanics of an oscillator with time dependent frequency with the well known difficulties arising from the ambiguity of vacuum state in a time dependent background. The black hole spacetime, of course, does not posses such a spatially homogeneous description. The Rindler spacetime does but, in this case, the homogeneous description is just the inertial coordinates in flat spacetime with no time dependence --- thereby making that description trivial.

Another issue which crops up in the study of de Sitter spacetime is the following. Neither the inflationary phase nor the currently accelerating phase of the universe is strictly de Sitter. While one can provide a fairly elegant mathematical description of quantum field theory in \textit{exact} de Sitter spacetime, many of these techniques will fail when the spacetime is only \textit{approximately} de Sitter. It is interesting to ask how much progress one can make in studying such (approximately de Sitter) spacetimes and how much of the results valid in exact de Sitter will continue to hold (in an approximate sense) in such spacetimes. For example, it is fairly complicated to work with an analogue of approximately static coordinate system when the universe is not strictly de Sitter.  Many of the techniques used to define the vacuum states in exactly de Sitter spacetimes will also be inapplicable when the manifold has no de Sitter symmetry.

It is thus clear that, given the special features possessed by de Sitter spacetime, one can approach the problem of quantum field theory in de Sitter spacetime from many different perspectives, not all of which will be easily generalizable to an approximately de Sitter spacetime. This motivates us to examine closely several aspects of quantum field theory in de Sitter spacetime delineating the properties which arise (in one way or the other) from the symmetry of de Sitter spacetime from those which are of more general nature. Such a study also reveals some significant differences between de Sitter spacetime in (1+1) dimension and de Sitter spacetime in (1+3) dimensions.

We will now briefly describe some of these issues which will be discussed in detail in the paper. As we mentioned earlier, there is a very standard procedure for obtaining the thermality of the horizon in the \emph{static} coordinate system. This procedure works for a very wide class of spacetimes and can, for example, handle Rindler, black hole and de Sitter spacetimes at one go. But when the cosmological spacetime is not exactly de Sitter, no static coordinate system will exist. One can still define an ``approximately static'' coordinate system but this proves to be difficult to handle mathematically. 

If one decides to work with a Friedmann coordinate system, then the mathematics simplifies considerably because we will be dealing with a quantum mechanical problem rather than quantum field theory. But conceptually, we now have to tackle the issue of defining the vacuum state in a time dependent background. This turns out to be reasonably straight forward in (1+1) dimension in which conformal invariance of a massless scalar field helps the analysis. But in (1+3) dimension it is not possible to have sensible limits for the mode functions in the infinite past if one works with the massless scalar field $\Phi(t, \mathbf{x})$ as the primary variable. The usual trick is to work instead with the variable $\chi (t, \mathbf{x}) \equiv a(t)\Phi(t, \mathbf{x})$ and define a vacuum state in the asymptotic past for $\chi(t, \mathbf{x})$. When $t\to -\infty$, $a(t) \to 0$ and it is the scaling out of this factor which allows us to define a sensible vacuum state in (1+3) dimensions. The resulting vacuum state is the well known Bunch-Davies vacuum \cite{bunchdavies}  for the de Sitter spacetime. Clearly, this depends on the behaviour of $a(t)$ and there is no natural analogue of this vacuum state for non-de Sitter spacetimes.

An alternative to the above procedure is to define a vacuum state at some fixed time $t=t_0$, say, by choosing the modes which behave as close to the positive frequency modes as possible at this instant. We will call this the \textit{co-moving vacuum} since it is based on the co-moving time coordinate of Friedmann spacetime. In general, this vacuum state differs from the Bunch-Davies vacuum but it has the advantage that the evolution of $a(t)$ for $t<t_0$ becomes irrelevant for its definition. It is, therefore, well suited to study spacetimes which are de Sitter at late times with deviations from de Sitter geometry in the early epochs.  

Once the vacuum state is defined, in the asymptotic past or at some other chosen moment, one could study the mixing of positive and negative frequency modes due to the time dependence of the background expansion. In particular, one would be interested in knowing whether the mixing leads to a thermal nature for the state at later times. It does happen in the case of (1+1) dimension but the spectrum is \textit{not} strictly Planckian in the case of (1+3) dimensions. There are some interesting peculiarities which arise in this context when we try to obtain thermality \emph{working entirely} in the Friedmann coordinates.

Finally, one can also study the inter-relationship between the mode functions defined in Friedmann coordinate system and those defined with static coordinates. This is an exercise in evaluating the Bogolioubov coefficients and we do find that one recovers standard Planck spectrum without any deviation. This allows us to establish a correspondence between the vacuum states defined using the two coordinate systems but --- since static coordinate systems do not exist for approximately de Sitter spacetimes --- the approach does not allow an easy generalization to more realistic cases.

In the last part of the paper we study the mode functions in approximately de Sitter spacetimes in Friedmann coordinates. We find an explicit solution to the wave equation, correct to the necessary order of approximation, and use it to describe the deviations from the exact de Sitter spacetime.  This approach is quite general and is capable of handling a wide variety of cases when the evolution is approximately de Sitter.

The plan of the paper is as follows:  We briefly review thermal aspects of  horizons in static coordinate system in \sec{sec:thermality}. In \sec{sec:fieldmodes}, we solve for the  modes of a massless scalar field in spatially flat de Sitter spacetime in (1+1) and (1+3) dimensions and define the Bunch-Davies and comoving vacuum states in the Friedmann coordinate patch. These modes evolve in time and the physical content of the modes at later times is determined by evaluating the mixing coefficients   in \sec{sec:particlecontent}, working entirely in the Friedmann coordinates. In \sec{sec:detectorresponse},  we study the response of a detector coupled with the field as a way to provide an operational meaning to the mixing coefficients. We next compare, in \sec{sec:static}, the mode functions defined in the Friedmann patch with those defined in the static patch of the de Sitter spacetime  to reproduce some standard results. Finally, we study the corresponding effects in the quasi-de Sitter geometry. We consider a small perturbation to the de Sitter metric  and develop the perturbative framework to find the corrections to the field modes and the corresponding power spectrum in the quasi-de Sitter case. This procedure is  illustrated by taking the model of the universe containing pressure-free matter and cosmological constant which behaves like quasi-de Sitter at late times. Section \ref{sec:conclusions} describes the conclusions.


\section{Thermality in static coordinates}
\label{sec:thermality}

To set the stage, we shall begin by briefly reviewing some well known results (see e.g., chapter 14 of Ref.~\cite{paddygrav} for more details) related to the temperature of horizons in static coordinates. Several spacetimes of interest including the Schwarzschild, de Sitter, Rindler, can be described by a line element of the form
\begin{equation}
\label{frmetric}
ds^2 = f(r)dt^2 - \frac{dr^2}{f(r)} - dL_\bot^2
\end{equation}
where $dL_\bot^2 $ is the transverse metric and  $f(r)$ vanishes at the horizon, $r = r_0$ with $f'(r_0) \equiv 2\kappa \neq 0$. Then, using a Taylor series expansion near the horizon, we can write $f\approx 2\kappa l$ where $l = (r - r_0)$ and the metric near the horizon takes the form:
\begin{equation}
ds^2 = 2\kappa l\, dt^2 - \frac{dr^2}{2\kappa l} - dL_\bot^2.
\end{equation}
In the case of Rindler spacetime, this is exact and $\kappa$ denotes the acceleration of the Rindler observer. In other cases, the metric reduces to this form close to the horizon with $\kappa$ denoting the surface gravity. 

This (Rindler) form of the metric makes it obvious that the singular behaviour of the metric near $l=0$ is a coordinate artefact. It is possible to introduce several, different, sets of coordinates which will cover the entire manifold without any pathology at the horizon. One such choice, $(T,\textbf{X})$, which we will call \textit{Kruskal-like coordinates} is obtained by the transformations:
\begin{equation}
\kappa X = e^{\kappa r_*} \cosh \kappa t;\hspace{7pt} \kappa T = e^{\kappa r_*}\sinh \kappa t; \hspace{10pt} r_* \equiv \int\frac{\dd r}{f(r)}
\end{equation}
which
lead to the  metric
\begin{equation}
ds^2 = \frac{f}{\kappa^2(X^2 - T^2)}(dT^2 - dX^2) + dL_\bot^2
\end{equation} 
that covers the full manifold. Here  $f$ should be treated as a function of $(T,X)$.  The horizon at $r=r_0$ is now  mapped to $T^2 = X^2$ but with the factor $f/(X^2-T^2)$ remaining finite at the horizon. 

It is now possible to show that the vacuum state of a quantum field defined on the $T=0$ hypersurface appears as a thermal state to observers confined on the right wedge $X>|T|$. This is most easily done by making an analytic continuation to the imaginary time coordinates by $T_E = -i T$ and $t_E = -it$. The time evolution of the system in terms of  $T_E$ will take the field configuration from $T_E = 0$ to $T_E\rightarrow\infty$ and will be governed by a global  Hamiltonian $H_{gl}$. 
One can equivalently describe the same evolution in terms of  $t_E$, which  behaves like an angular coordinate from $t_E=0$ to $t_E =  2\pi/\kappa$  when we use  the  Hamiltonian $H_{st}$ which determines time evolution in the static time coordinates. The  entire upper half-plane $T>0$ can be covered in two completely different ways: either in terms of the evolution in $T_E$ or in terms of the evolution in $t_E$. In $(T_E,X)$ coordinates, we vary $X$ in the range $(-\infty, \infty)$ for each $T_E$ and vary $T_E$ in the range $(0,\infty)$. In $(t_E,x)$ coordinates, $x$ varies in the range $(0,\infty)$ for each $t_E$ which 
varies  in the range $(0,\pi/\kappa)$ like an angular variable. This  allows us to prove, using standard path integral techniques \cite{lee,paddygrav}  that,
\begin{equation}
\langle\mathrm{vac}|\phi_L,\phi_R\rangle \propto \langle\phi_L|e^{-\pi H_{st}/\kappa}|\phi_R\rangle.
\end{equation} 
where $\phi_L$ and $\phi_R$ are the field configurations in left and right parts of the plane on the $T=0$ hypersurface. 
One can find the density matrix for observations confined to the right wedge by tracing out the field configuration $\phi_L$ on the left wedge. This computation gives:
\begin{equation}
\rho (\phi_R,\phi_R') = \frac{\langle\phi_R|e^{-2\pi H_{st}/\kappa}|\phi_R'\rangle}{Tr(e^{-2\pi H_{st}/\kappa})}
\end{equation}
which is thermal with the temperature $\beta^{-1} = \kappa/2\pi$.
Thus, the vacuum state of the field defined on the $T=0$ hypersurface leads to a thermal density matrix with temperature $\kappa/2\pi$ as far as static observers in the right hand wedge are concerned. 

In the case de Sitter the  metrics in the static and Kruskal-like coordinates
\begin{eqnarray}
ds^2&=& (1 - H^2 r^2)dt^2 - \frac{dr^2}{(1- H^2 r^2)} - dL_\bot^2 \nonumber\\
&=&  \frac{4}{[H^2(X^2-T^2)+1]^2} (dT^2 - dX^2) - dL_\bot^2
\end{eqnarray} 
are connected by the coordinate transformations:
\begin{equation}
X = \frac{1}{H}\left(\frac{1+Hr}{1-Hr}\right)^{1/2} \cosh H t, \qquad
T = \frac{1}{H}\left(\frac{1+Hr}{1-Hr}\right)^{1/2} \sinh H t;
\end{equation} 
From the form of the metric in in the two coordinate systems, it is obvious that 
the $(T, \mathbf{X})$
coordinate system is not static because the metric  depends on $T$.  On the other hand, the coordinate system $(t,\mathbf{x})$
which covers the right wedge has a static time coordinate $t$. 
It is well known that defining a vacuum state in a time dependent background is non-trivial and often ambiguous. In the above analysis we have chosen to define a vacuum state at a \emph{particular} space-like hypersurface $T=0$ and examine its properties in terms of the static coordinates. A different definition for the vacuum state, in general, will lead to a different description in static coordinates. We will see that similar issues arise later on when we study de Sitter universe in the Friedmann coordinates as well.

An alternative procedure to determine the thermal nature of the horizon is  based on the calculation of relevant Bogoliubov coefficients.  Since we have two coordinate systems --- Kruskal-like and static --- covering part of the manifold, one can obtain, in principle, the relation between the field modes which are natural to these coordinate systems and compute the Bogoliubov coefficients between them. 
Let the field modes be given by some functions $\phi(T,X)$ and $\chi(t,r)$ in the Kruskal-like and static coordinates respectively in the region of the manifold where both are well defined. (For simplicity, we have ignored the dependence on  the transverse coordinates which play no role in the discussion, as we shall see.)  
It is often not possible to obtain closed expressions for the field modes due to mathematical complexity. However, it is possible to evaluate the Bogoliubov coefficients using a simple trick: Since the Bogoliubov coefficients that relate the two sets of field modes  are independent of the hypersurface which is used to evaluate the Klein-Gordon inner product, 
we can choose this hypersurface to be arbitrarily close to the horizon.  The field equations reduce to a \emph{two-dimensional} wave equation near the horizon making the dependence in the transverse coordinates (and the mass of the field) irrelevant. Conformal invariance then allows us to determine the field modes near the future horizon which take the form of plane waves in the relevant coordinates. That is, 
\begin{equation}
\chi_\omega (t,r) = \frac{1}{\sqrt{2\omega}}e^{-i\omega u}
\end{equation}
and 
\begin{equation}
\phi_k(T,X) = \frac{1}{\sqrt{2k}}e^{-ik U}.
\end{equation}
where $u = t - r^*$ and $U = T- X$ respectively. These are related by $\kappa U = - e^{-\kappa u}$ 
which signifies an exponential redshift near the horizon. As is well known, the relevant Bogoliubov coefficient (which we will have the occasion to evaluate explicitly later on) will now lead to a thermal spectrum of particles. 

This discussion is, of course, applicable to the de Sitter universe described by the metric in \eq{frmetric} with $f(r) = 1 - H^2r^2$ and will lead to a temperature $H/2\pi$. More precisely, if we introduce Kruskal-like coordinates in the de Sitter manifold and define a vacuum state on the $T=0$ hypersurface, then such a vacuum state will lead to a density matrix with temperature $H/2\pi$ for the observers confined to the region $r<H^{-1}$. Once again, it should be stressed that the de Sitter metric in the Kruskal-like coordinates is \emph{not} static and the vacuum state is defined using the $T=0$ hypersurface.  
   
This analysis is completely in tune with what could be done in black hole spacetimes as well as in the case of Rindler spacetime.  But in the case of de Sitter we have an \emph{alternative coordinate system} available to us, viz.., the standard Friedmann coordinate system.  This allows us to study the dynamics of a quantum field \emph{entirely} in the Friedmann coordinate system and explore whether we can recover the thermality of the horizons and other features. In such a study we necessarily have to work with a time dependent background but --- as we have emphasized above --- this is implicit \textit{even when we use Kruskal-like coordinates} and relate them to static coordinates.  We can, therefore, adopt a similar strategy in the Friedmann coordinate system by defining a vacuum state at some suitable hypersurface and studying its particle content as the evolution proceeds.
   
This approach has one extra advantage. The static coordinate system exists only for the \textit{exact} de Sitter universe. When there are deviations from de Sitter nature, we can still describe the universe in a very natural fashion using Friedmann coordinate system. But in this case, we will not have the luxury of an alternative static coordinate system to describe the physics. Therefore, a formalism which addresses issues like thermality working entirely in Friedmann coordinate system, without using any of the symmetries of the de Sitter universe, is well suited for the study of near de Sitter geometry.
We will find that obtaining thermal nature  of the horizon working entirely in Friedmann coordinates is --- surprisingly --- not an easy task. In fact we could \emph{not} find any previous work in published literature which discusses  quantum field theory in de Sitter spacetime from such an approach and obtains thermal nature of the horizon. We shall now turn to this study.


\section{Massless scalar field modes in de Sitter spacetime}
\label{sec:fieldmodes}

Throughout the paper, we will confine ourselves to massless, minimally coupled scalar field in de Sitter spacetime. The action for the field $\Phi(t,\textbf{x})$ is given by:  
\begin{equation}
S[\Phi] = \frac{1}{2} \int \dd^nx \, \sqrt{-g}\,\partial_a\Phi\partial^a\Phi
\end{equation}
It turns out that the dynamics is somewhat different in (1+1) dimensional spacetime compared to (1+3) dimensional spacetime. We will first study  the  behaviour in $(1+1)$ and then follow the same procedure for the  $(1+3)$ case. This will bring out the similarities and some curious differences between the two cases. 

\subsection{Field modes in $\mathbf{dS}_2$ spacetime}
\label{subsec:fieldmodesds2}

We will describe the de Sitter spacetime in Friedmann coordinates with $k=0$. Then, the (1+1) dimensional metric is given by, 
\begin{equation}
ds^2 = dt^2 - a^2(t) dx^2
\end{equation}
and
the field equation reads,
\begin{equation}
\partial^2_t\Phi + \frac{\dot{a}}{a}\,\partial_t\Phi - \frac{1}{a^2}\,\partial^2_\textbf{x}\,\Phi=0.
\label{wave}
\end{equation}
We decompose $\Phi$ in terms of a complete set of orthonormal functions $f_\textbf{k}$ in the form
\begin{equation}
\label{phiexp}
\Phi(t,x)=\int_{-\infty}^{\infty} \frac{dk}{2\pi}\ \left[\hat{a}_k f_k + \hat{a}^\dagger_k f^*_k\right].
\end{equation}
Spatial homogeneity allows us to separate out the $x$ dependence and write:
\begin{equation}
f_k(x,t)= e^{ikx}\, \psi_{|k|} (t)
\end{equation}
 Substituting in \eq{wave} and solving the resulting equation, we find that
\begin{equation}
\psi_k(t) = \mathcal{A}_k s_k(t) + \mathcal{B}_k s_k^*(t);\hspace{7pt}s_k(t) = \frac{1}{\sqrt{2k}}\exp \left(-ik\int\frac{\dd t}{a(t)}\right)
\label{modechoice}
\end{equation}
(The $k$ in these expressions actually stand for $|k|$; we will not explicitly show the modulus sign hereafter for notational simplicity.)
The result is obvious from the fact that in (1+1) dimension the scalar field action is conformally invariant and any Friedmann spacetime is conformally flat with the conformal time coordinate $\eta$ defined through $d\eta = dt/a(t)$.  For  $\mathrm{dS}_2$ with the scale factor $a(t) = e^{Ht}$, the  solution is,
\begin{equation}
\label{s1pl1}
s_k(t) = \frac{1}{\sqrt{2k}}\exp\left[-\frac{ik}{H}\left(1- e^{-Ht}\right)\right]
\end{equation}
where the phase ensures that, in the $H\rightarrow0$ limit, the mode function reduces to standard flat spacetime modes. The constants $\mathcal{A}_k$ and $\mathcal{B}_k$ in \eq{modechoice} are determined using the appropriate boundary conditions and thus decide the choice of the  vacuum  state for the field. For example, when $H\to 0$, the choice $\mathcal{A}_k = 1$ and $\mathcal{B}_k = 0$ gives the positive frequency mode and selects the standard inertial vacuum for the flat spacetime. 

In the presence of an expanding background, it is difficult to define a unique choice for the vacuum and we need to study different choices and their physical properties.
One possible choice would be to define the vacuum state at  the asymptotic past by choosing the field modes such that they reduce to positive frequency modes in this limit. 
It is, however, clear that the mode function in \eq{s1pl1} does \textit{not} have a well defined phase when $t\rightarrow -\infty$.
(This is related to the fact that in the asymptotic past $a\to 0$.) The usual procedure adopted in the literature to circumvent this problem is to abandon the idea of defining a vacuum state using the $t$ coordinate and instead use the conformal time $\eta$ which, for the de Sitter universe, can be taken to be  $\eta \equiv (1- e^{-Ht})/H$. (The integration constant is chosen to give the correct limit of $\eta \to t$ when $H\to 0$.) Then our mode function in \eq{s1pl1}
\begin{equation}
s_k(\eta) = \frac{1}{\sqrt{2k}}e^{-i k\eta}
\end{equation}
is indeed a positive frequency solution with respect to $\eta$ (at all times) and therefore the choice $\mathcal{A}_k = 1$ and $\mathcal{B}_k = 0$ gives a natural choice for the vacuum. This is the conventional Bunch-Davies vacuum defined with respect to conformal time
by the choice of mode functions
\begin{equation}
\psi_{k}^{(BD)} (t)= \frac{1}{\sqrt{2k}} \exp\left[-\frac{ik}{H}\left(1- e^{-Ht}\right)\right]
\end{equation}

While the Bunch-Davies vacuum is the preferred choice in the literature, it is clear that it is more in tune with the conformal time coordinate $\eta$ than with the co-moving time coordinate $t$. In the Friedmann metric, the co-moving time $t$ has a direct physical significance as the proper time of the co-moving, geodesic clocks. This motivates us to look at the possibility of defining a \textit{co-moving vacuum} with mode functions which behave as close as possible to the positive frequency modes with respect to co-moving time coordinate $t$. We can take a cue from the discussion in the last section where we saw that, even in the Kruskal-like coordinates for the de Sitter spacetime, the metric is time dependent and the vacuum state is defined on a \textit{particular} hypersurface $T=0$. 
In a similar fashion, we can choose the modes in \eq{modechoice} by demanding that at some time $t=t_0$ they behave like positive frequency modes. Because of the time translational invariance, we can take $t_0=0$, without the loss of generality, as long as $t_0$ is finite. That is, we impose the conditions:
\begin{equation}
\psi_{k} (0) = \frac{1}{\sqrt{2k}}e^{-ikt}|_{t=0}\,\,;\hspace{7pt}\dot{\psi}_{k}(0) = \frac{-ik}{\sqrt{2k}}\,e^{-ikt}|_{t=0}.
\label{cond}
\end{equation}
(The same physics is obtained if we take $t_0\neq 0$ with the replacement of $k$ by $k e^{-Ht_0}$ which ensures that $k$ is the co-moving wave number defined at $t_0=0$).
These conditions imply that at $t=t_0(=0)$ the mode function and its derivative behave like a positive frequency mode.

We can now determine the coefficients $\mathcal{A}_k$ and $\mathcal{B}_k$ using this condition and --- somewhat curiously --- we will \emph{again find} that  $\mathcal{A}_k = 1$ and $\mathcal{B}_k = 0$. That is, the mode function   $\psi_k^{(CM)} (t)$, evolved from the \emph{co-moving} vacuum defined at $t = t_0(=0)$ is same as Bunch-Davies state $\psi_k^{(BD)} (t)$  defined earlier in $\mathrm{dS}_2$. 
This result is independent of the choice for $t_0$ thereby showing that the Bunch-Davies vacuum can \textit{also} be interpreted as a co-moving vacuum state defined using the conditions in \eq{cond}. 

We will see later, this equivalence is a special feature of (1+1) dimension and does not hold in (1+3) dimensions where the co-moving and Bunch-Davies vacua are different.


\subsection{Field modes in $\mathbf{dS}_4$ spacetime}
\label{subsec:fieldmodesds4}
We shall now follow the same procedure as above in the $(1+3)$ dimensions. The metric is now given by
\begin{equation}
ds^2 = dt^2-\exp(2Ht) d\textbf{x}^2
\end{equation}
where $H$ is the Hubble constant and sets the only length-scale (or time-scale) in the problem to be $1/H$. The field equation for $\Phi(t,\textbf{x})$ in this metric reads:
\begin{equation}
\partial^2_t\Phi + 3H\partial_t\Phi - \exp(-2Ht)\partial^2_\textbf{x}\,\Phi=0
\end{equation}
As usual we expand the field in terms of a complete set of orthonormal functions $f_\textbf{k}$
and write:
\begin{equation}
\label{phiexp1}
\Phi(\textbf{x},t)=\int\frac{d^3k}{(2\pi)^3} \left\{\hat{a}_\textbf{k} f_\textbf{k}(t,\textbf{x}) + \hat{a}^\dagger_\textbf{k} f^*_\textbf{k}(t,\textbf{x})\right\}
\end{equation}
where spatial homogeneity allows us to express the field modes in the form:
\begin{equation}
f_\textbf{k}(t,\textbf{x})= e^{i\,\textbf{k}\cdot \textbf{x}}\, \psi_k(t)
\end{equation}
where $k=|\textbf{k}|$. The equation in $\psi_k(t)$ then becomes,
\begin{equation}
\ddot{\psi}_k+3H\dot{\psi}_k+\exp(-2Ht)k^2\psi_k=0.
\end{equation}
with the solution,
\begin{equation}
\label{psi_0}
\psi_k(t)= \mathcal{A}_k s_k (t) +  \mathcal{B}_k s_k^* (t)
\end{equation}
where
\begin{equation}
\label{s}
s_k (t) = \frac{1}{\sqrt{2k}}\exp\left[-\frac{ik}{H}\left(1- e^{-Ht}\right)\right]\left(\frac{iH}{k} + e^{-Ht}\right)
\end{equation}
(Because of the existence of  $\mathcal{A}_k$ and $\mathcal{B}_k$, the normalization of $s_k$ is not unique; we choose it in such a way that, when $\mathcal{A}_k=1$ and $\mathcal{B}_k=0$, the functions $s_k$ satisfy the standard orthonormality conditions with respect to the Klein-Gordon inner product.)
Again, the constants $\mathcal{A}_k$ and $\mathcal{B}_k$ are to be determined using the appropriate boundary conditions which makes a choice for the vacuum state for the field. In (1+3) dimensions also, we see that when $H\to 0$ the choice $\mathcal{A}_k= 1$ and $\mathcal{B}_k = 0$  leads to the standard  positive frequency mode in flat spacetime and  selects the inertial vacuum. Our interest is to explore the different choices in the presence of expanding background. 

As in $\mathrm{dS}_2$ case,  let us first study the behaviour of the modes in  the asymptotic past. We see that, in the $t\rightarrow -\infty$ limit, the expression in \eq{s} goes to:
\begin{equation}
s_k(t) \rightarrow  \frac{1}{\sqrt{2k}}\exp\left(\frac{ik}{H}e^{-Ht}\right) e^{-Ht}.
\end{equation}
This does not have a well-defined limit and hence cannot be used to define a vacuum state for the field. In this respect, both (1+1) and (1+3) dimensional results are similar.

We found that, in the (1+1) dimensional case we could use the conformal time coordinate $\eta$ to define a natural vacuum state in the asymptotic past. In the present case, however, the situation is different. In terms of the conformal time $\eta$, the mode function becomes (in the asymptotic past): 
\begin{equation}
\label{setalimit}
s_k(\eta) \rightarrow \frac{1}{a(\eta)}\frac{e^{-i k\eta}}{\sqrt{2k}}
\end{equation}
and we see now the crucial difference  from the $(1+1)$ dimensional case. There is an extra $a(\eta)$ in this case which prevents us from treating it as the standard positive frequency mode. 

The result also suggests a possible way-out which is usually adopted in the literature. Instead of quantising $\Phi$ we may choose to quantise $\bar\Phi\equiv a(t)\Phi$. This is a (time dependent) point transformation of the dynamical variable which is permissible in the classical description.
We then see that the choice $\mathcal{A}_k = 1$ and $\mathcal{B}_k = 0$ will give the modes $\exp(-ik\eta)$ which has the standard form for $\bar\Phi$, when treated as a quantum field. This is the usual procedure in the literature and this choice leads to the conventional Bunch-Davies vacuum. But note the the situation was \textit{not} as straightforward as in the case of $(1+1)$ dimensions and we needed to remove a factor $a(t)$ to define the vacuum state in (1+3) dimensions. 
 
The difference is more acute when we try to define a co-moving vacuum. As in the case of $(1+1)$ dimension one can define the co-moving vacuum by imposing the conditions given in \eq{cond} and thus determining $\mathcal{A}_k$ and $\mathcal{B}_k$.   Because of the time translation invariance, we can again define the co-moving vacuum at at $t = 0$ and the result for any other time, $t_0$ can be obtained by a finite shift. Hence the  conditions we impose on the modes are
\begin{equation}
\psi_{k} (0) = \frac{1}{\sqrt{2k}}e^{-ikt}|_{t=0}\,\,;\hspace{7pt}\dot{\psi}_{k}(0) = \frac{-ik}{\sqrt{2k}}\,e^{-ikt}|_{t=0}.
\end{equation}
These allow us to determine the constants $\mathcal{A}_k$ and $\mathcal{B}_k$ as:
\begin{equation}
\label{ABcm}
\mathcal{A}_k=\frac{H+2ik}{2ik};\hspace{5pt} \mathcal{B}_k=\frac{H}{2ik}
\end{equation}
which define the mode function, $\psi_k^{(CM)} (t)$, evolved from the \emph{co-moving} vacuum choice defined at $t = 0$.

When we did this in (1+1) dimension, we found that $\mathcal{A}_k = 1$ and $\mathcal{B}_k = 0$ - instead of the expressions in \eq{ABcm} ---  thereby showing the equivalence of co-moving and Bunch-Davies vacuum. But in (1+3) dimensions we get a different result, viz. that the co-moving vacuum is \textit{different} from the Bunch-Davies vacuum. The difference can be traced, algebraically, to the existence of the $a(\eta)$ factor in \eq{setalimit}.

To summarise, we can define the  vacuum states by imposing suitable boundary conditions on the mode functions and thus determining the constants $\mathcal{A}_k$ and $\mathcal{B}_k$. If we work in the asymptotic past, then one can choose the modes to be $\exp(-ik\eta)$ in (1+1) dimension, thereby defining the Bunch-Davies vacuum. In (1+3) dimensions this is not possible with the original scalar field. But if we work with $a(t)\Phi$, instead of $\Phi$, one can again define the modes such that they behave as  $\exp(-ik\eta)$ in the asymptotic past. Alternatively, one can attempt to define a co-moving vacuum by imposing the condition that the modes must behave as close to positive frequency solutions as possible, with respect to the co-moving time coordinate $t$, at some time $t=t_0$. Because of time translation invariance, we can choose $t_0=0$ without loss of generality. We then find that, in (1+1) dimension, the co-moving vacuum is equivalent to the Bunch-Davies vacuum. But in (1+3) dimensions, these two mode functions (and hence the vacua are different). We shall now explore the properties of these vacuum states.


\section{Evolution and mixing coefficients at later times}
\label{sec:particlecontent}

The Bunch-Davies and the co-moving vacua are defined by the condition that the mode function is purely positive frequency at a given moment of time $t=t_0$. In the case of Bunch-Davies vacuum, this  is done in the asymptotic past ($t_0\to-\infty$) while in the case of co-moving vacuum we choose this to be $t_0=0$. Once this initial condition is set, expansion of the universe will evolve the mode functions to a mixture of positive and negative frequency modes, with respect to the co-moving time coordinate, at any later time. This mixing can be analysed in terms of two mixing coefficients, $\alpha_\nu$ and $\beta_\nu$ in the expansion:
\begin{equation}
\psi_k (t) = \int_0^{\infty}\,\frac{\dd\nu}{2\pi} \left(\alpha_\nu\,e^{-i\nu t} + \beta_\nu\,e^{i\nu t} \right)
\label{mix}
\end{equation}
It is slightly more convenient to let the frequency vary over both positive and negative values and write:
\begin{equation}
\label{eq:ft}
\psi_k(t) = \int_{-\infty}^{\infty} \frac{\dd\nu}{2\pi}\,f(\nu)\,e^{-i\nu t}
\end{equation}
so that 
\begin{equation}\label{alphabeta}
\alpha_\nu = f(\nu),\hspace{5pt}\beta_\nu = f(-\nu); \hspace{7pt}\nu > 0
\end{equation}
The mixing coefficients defined by \eq{mix} are similar to Bogoliubov coefficients but \textit{not} the same. We stress that in \eq{mix} $\psi_k (t)$ is expanded in terms of the complete set of orthonormal functions $\exp(\pm i\nu t)$ which are \textit{not} the solutions to scalar field wave equation in the de Sitter background. Physically, one can think of these functions $\exp(\pm i\nu t)$ as defining the instantaneous positive and negative frequency mode functions with respect to the co-moving time. But as we shall see, these mixing coefficients have interesting properties and in fact play a direct role in the response of detectors. We shall say more about it later on.

The task of determining the mixing coefficients is thus reduced to calculating the the Fourier transform of $\psi_k(t)$,
\begin{equation}
f(\nu) = \int_{-\infty}^\infty \mathrm{d}t\,\,e^{i\nu t}\,\psi_k(t) .
\end{equation}
Often we will be interested in  $|\alpha_\nu|^2$ and $|\beta_\nu|^2$ which can be obtained from the power spectrum $|f(\nu)|^2$. We shall now compute these for the different cases.


\subsection{Mixing coefficients in $\mathbf{dS}_2$}
\label{subsec:plecontentds2}

We will begin with the (1+1) dimensional case for which the modes are
\begin{equation}
\psi_{k}(t)= \frac{1}{\sqrt{2k}} \exp\left[-\frac{ik}{H}\left(1- e^{-Ht}\right)\right]
\label{sol1}
\end{equation}
So, the Fourier transform is:
\begin{align}
f(\nu) &= \frac{e^{-ik/H}}{\sqrt{2k}}\int_{-\infty}^\infty \mathrm{d}t\,\,e^{i\nu t}\,e^{ik/H\, e^{-Ht}}\nonumber\\
          &= \frac{e^{-ik/H}}{\sqrt{2k}}\left(\frac{1}{H}\right)\left(\frac{k}{H}\right)^{i\nu/H}\Gamma\left(-\frac{i\nu}{H}\right) e^{\pi\nu/2H}.
\end{align}
Similarly,
\begin{align}
f(-\nu) = \frac{e^{ik/H}}{\sqrt{2k}}\left(\frac{1}{H}\right)\left(\frac{k}{H}\right)^{i\nu/H}\Gamma\left(-\frac{i\nu}{H}\right) e^{-\pi\nu/2H}.
\end{align}
so that the modulus square of the coefficients of mixing in \eq{alphabeta} are given by :
\begin{align}
|\alpha_\nu|^2 = \frac{1}{2k\nu}\frac{\beta e^{\beta\nu}}{e^{\beta\nu} - 1};\quad
|\beta_\nu|^2 = \frac{1}{2k\nu}\frac{\beta}{e^{\beta\nu} - 1};\quad\beta = \frac{2\pi}{H}.
\end{align}
That is the power spectrum per logarithmic band  at negative frequencies (given by $|\beta_\nu|^2$) is Planckian at temperature, $H/2\pi$. At first sight this might look like the familiar result, well known in literature. However, there are some peculiar features which need to be commented on. 

Note that we have started with a solution to the wave equation in de Sitter background (given by \eq{sol1}) and expanded it using the complete set of functions $\exp(\pm i \nu t)$. \textit{These functions have no ``legality'' in the de Sitter spacetime} since they are not the solutions of the wave equation. We could have, for example, used any other complete set of orthonormal functions in place of $\exp(\pm i \nu t)$ and could have defined the mixing coefficients through an equation like \eq{mix}.  The two properties which  favour  our choice are: (a) they were precisely the mode functions used to define the co-moving vacuum and (b) they are instantaneous, monochromatic, plane waves with respect to the co-moving time $t$.  It is therefore interesting that the overlap between positive and negative frequencies in such an expansion gives rise to the thermal spectrum. 

If we had used the conformal time instead of co-moving time, then the result would have been very different ---  and very trivial. In terms of the conformal time, the modes are just $\exp(\pm ik \eta)$ at all $\eta$ and there is \textit{no} mixing of the positive and negative frequencies defined with respect to $\eta$. So if we had defined another set of mixing coefficients with an equation like \eq{mix} but with conformal time $\eta$, then we would have got the trivial result $\beta_\nu =0$. So if we define the vacuum state with respect to conformal time and work entirely in terms of conformal time we will see no trace of thermal spectrum in the de Sitter universe. This is, of course, obvious from the fact that the metric is conformally flat in $(\eta, x)$ coordinates and the scalar field theory is conformally invariant in (1+1) dimension; so we are back to the evolution of inertial vacuum in flat spacetime. On the other hand, the modes undergo exponential redshift when frequencies are defined with respect to co-moving time and --- as we had already mentioned --- the exponentially redshifted wave will lead to a thermal mixing coefficient. We will next see that the situation is somewhat different in the (1+3) dimensional case.

\subsection{Mixing coefficients in $\mathbf{dS}_4$}
\label{subsec:plecontentds4}

In this case, which we want to study in detail, it is convenient to work with a general mode function, having arbitrary coefficients $\mathcal{A}_k$ and $\mathcal{B}_k$.  From \eq{psi_0}, we have, 
\begin{equation}
\psi_k(t) = \mathcal{A}_k s_k (t) +  \mathcal{B}_k s_k^* (t) =  \psi_k^{(1)}(t) + \psi_k^{(2)}(t)
\end{equation}
where
\begin{equation}
s_k (t) = \frac{1}{\sqrt{2k}}\exp\left[-\frac{ik}{H}\left(1- e^{-Ht}\right)\right]\left(\frac{iH}{k} + e^{-Ht}\right)
\end{equation}
Recall that taking $\mathcal{A}_k = 1$ and $\mathcal{B}_k = 0$ gives Bunch-Davies state and for the co-moving state the corresponding values are provided by \eq{ABcm}. 
We again choose the normalization such that $f_\textbf{k} \equiv  s_k \exp(i\mathbf{k\cdot x})$ satisfies the standard orthonormality conditions
$(f_\textbf{k},f_{\textbf{k}'}) = \delta^3(\textbf{k} - \textbf{k}')$ with respect to the Klein-Gordon inner product: 
\begin{equation}
(\Phi_1,\Phi_2) \equiv -i\int_t\dd^3x\,a^3(t)[\Phi_1\partial_t\Phi_2^* - \Phi_2^*\partial_t\Phi_1]
\end{equation}
where the integral is evaluated over a hypersurface of constant $t$. 

Some amount  of algebra yields the following results for the Fourier transform of the respective parts, $\psi_k^{(1)}(t)$ and $\psi_k^{(2)}(t)$:
\begin{align}
f^{(1)}(\nu) &= \mathcal{A}_k\frac{2e^{- ik/H}e^{\pi\nu/2H}}{(2k)^{3/2}} \left(\frac{k}{H}\right)^{\frac{i\nu}{H}}\Gamma\left(-\frac{i\nu}{H}\right)\left(i+\frac{\nu}{H}\right)\nonumber\\
f^{(2)}(\nu) &= \mathcal{B}_k\frac{2e^{ik/H}e^{-\pi\nu/2H}}{(2k)^{3/2}}\left(\frac{k}{H}\right)^{\frac{i\nu}{H}}\Gamma\left(-\frac{i\nu}{H}\right)\left(-i-\frac{\nu}{H}\right).
\end{align}
Taking the square of the modulus of the above expressions we get, 
\begin{align}
\nu |f^{(1)}(\nu) |^2 &=\frac{H^2}{2k^3}|\mathcal{A}_k|^2\frac{\beta e^{\beta \nu}}{e^{\beta \nu}-1}\left(1+\frac{\nu^2}{H^2}\right)\nonumber\\
\nu |f^{(2)}(\nu) |^2 &= \frac{H^2}{2k^3}|\mathcal{B}_k|^2\frac{\beta}{e^{\beta \nu}-1}\left(1+\frac{\nu^2}{H^2}\right)
\end{align}
where $\beta = 2\pi/H$. For negative frequencies, the forms of power spectrum are:
\begin{align}
\nu |f^{(1)}(-\nu) |^2 &=\frac{H^2}{2k^3}|\mathcal{A}_k|^2\frac{\beta }{e^{\beta \nu}-1}\left(1+\frac{\nu^2}{H^2}\right)\nonumber\\
\nu |f^{(2)}(-\nu) |^2 &= \frac{H^2}{2k^3}|\mathcal{B}_k|^2\frac{\beta e^{\beta \nu}}{e^{\beta \nu}-1}\left(1+\frac{\nu^2}{H^2}\right)
\end{align} 
Let us first consider the  Bunch-Davies state, with $\mathcal{A}_k=1$ and $\mathcal{B}_k=0$. This gives $\alpha_\nu = f^{(1)}(\nu)$ and $\beta_\nu = f^{(1)}(-\nu)$ so that we find:
\begin{equation}
|\alpha_\nu|^2 =\frac{H^2}{2k^3\nu}\frac{\beta e^{\beta \nu}}{e^{\beta \nu}-1}\left(1+\frac{\nu^2}{H^2}\right)
\end{equation}
\begin{equation}
|\beta_\nu|^2 = \frac{H^2}{2k^3\nu}\frac{\beta}{e^{\beta \nu}-1}\left(1+\frac{\nu^2}{H^2}\right)
\end{equation}
In contrast to the (1+1) dimensional case, these are not thermal, due to the extra factor 
$(1+\nu^2/H^2)$. So the expansion of the universe leads to a mixing of positive and negative frequencies but the resulting mixing coefficients do not have a thermal form. It should be noted, however, that the ratio of the mixing coefficients is
\begin{equation}
\frac{|\beta_\nu|^2}{|\alpha_\nu|^2} = \frac{|f^{(1)}(-\nu) |^2}{|f^{(1)}(\nu) |^2} = e^{-\beta\nu}
\label{ratio}
\end{equation}
When the field $\psi_k(t)$ couples linearly to a detector, the rate of upward and downward transitions between any two levels of the detector will be determined by the mixing coefficients. Therefore, when the condition in \eq{ratio} holds, one is led to
a level population in the detector at  thermal  equilibrium with the temperature $\beta^{-1} = H/2\pi$.
Any multiplicative function $h(\nu^2)$ with $\alpha_\nu$ and $\beta_\nu$ drops off in the ratio. 
[Usually, one works with \emph{Bogoliubov} coefficients which satisfy the constraint $|\alpha|^2 - |\beta|^2 =1$; in that case, if \eq{ratio} holds, then $|\beta|^2$ must be thermal. In the case of \emph{mixing} coefficients we have defined, the condition $|\alpha|^2 - |\beta|^2 =1$ does \textit{not} hold which allows extra factors like $(1+\nu^2/H^2)$.]

Let us next consider the co-moving vacuum which holds more surprises. We now require the square of the modulus of complete $f(-\nu)$ which is combination of individual quantities, $|f^{(1)}(-\nu)|^2$, $|f^{(2)}(-\nu)|^2$ evaluated above and a cross-term given by, 
\begin{equation}
2\nu |f^{(1)}(-\nu) ||f^{(2)}(-\nu) |\cos\theta =\frac{H^2}{k^3}|\mathcal{A}_\textbf{k}||\mathcal{B}_\textbf{k}|\left(1+\frac{\nu^2}{H^2}\right)\beta \frac{e^{\beta \nu/2}}{e^{\beta \nu}-1} \cos\theta
\end{equation}                                                                                                                                                                                                                                                                                 where
\begin{align}
\theta = \arg (f^{(1)},f^{(2)}). 
\end{align}
The complete expression becomes:
\begin{align}
\nu |f(-\nu) |^2 =  \frac{H^2\beta}{2k^3}&\left(1+\frac{\nu^2}{H^2}\right)\nonumber\\
&\left[\left(|\mathcal{A}_k|^2 + |\mathcal{B}_k|^2e^{\beta\nu}\right)  N + 2|\mathcal{A}_k||\mathcal{B}_k|\sqrt{N(N+1)}\cos \theta\right]
\label{res1}
\end{align}
where
\begin{equation}
N = \frac{1}{e^{\beta \nu} -1}
\end{equation}
is the Planckian factor and $\mathcal{A}_k, \mathcal{B}_k$ given by \eq{ABcm}. This  result shows that for a massless scalar field prepared in the \emph{co-moving} vacuum state, we obtain an expression having the Planckian factor with the temperature $H/2\pi$. 
In addition, we obtain an interference term involving $\sqrt{N(N+1)}$ which can be thought of as the fluctuation in the occupation number in thermal equilibrium. This factor has been noticed  earlier~\cite{supritsrini} in the case of horizon thermodynamics though no clear physical explanation is available. As far as we know, this has not been noticed earlier in the case of de Sitter spacetime in any context.

\section{Detector response in de Sitter spacetime}
\label{sec:detectorresponse}

The mixing coefficients defined through \eq{mix} are directly related to the response of a co-moving geodesic detector in Friedmann universe. Since the clock carried by such a detector will measure the co-moving time $t$, the rate of transition between the levels of the detector will involve the factors $\exp(\pm i t \Delta E)$ where $\Delta E$ is the energy difference between the two levels.  This gives an operational meaning to the mixing coefficients  and we will show that the response of a \textit{co-moving}, geodesic detector shows features very similar to what we obtained in the last section.

Consider a stationary detector, located at the spatial origin in a  de Sitter spacetime and  coupled to the massless scalar field by monopole interaction.  The amplitude for 
excitation of this detector  during the time interval $(-T, + T)$ due to its interaction with the scalar field can be computed, in  first order perturbation theory as:
\begin{equation}
\mathscr{A}_\textbf{k} = \mathcal{M} \int _{-T} ^T d\tau\, e^{i\nu\,\tau}\langle 1_\textbf{k} \vert \Phi (x[\tau]) \vert 0 \rangle
\label{amp}
\end{equation}
where $\mathcal{M} = i\lambda \langle E \vert \hat{m}(0)\vert E_0 \rangle$ is amplitude of transition in the internal levels of the detector with $\lambda$ as the coupling constant and $\hat{m}(0)$ is the detector's monopole operator. (In the above expression, we are confining our attention to a final field state containing a particle with a specified momentum $\mathbf{k}$. The total excitation probability for the detector is obtained by integrating $|\mathscr{A}_\textbf{k}|^2$ over all $\mathbf{k}$.) The detector interacts with the field only during the period $-T$ to $T$ and $x^a(\tau) = x^a(t) = (t,0,0,0) $ is the trajectory of the detector.  Expanding $\Phi(x[\tau])$  as in \eq{phiexp}, we find that the only term that survives in the $T\to \infty$ limit is the negative frequency term. The amplitude arising from this term is given by,
\begin{align}
\mathscr{A}_\textbf{k} =  \mathcal{M}\int^T _{-T}dt\,e^{i\nu t}\left[\frac{\mathcal{A}_k^*}{\sqrt{2k}}\right.&e^{ik/H} \left(-\frac{iH}{k} + e^{-Ht}\right)e^{-ik/H\,e^{-Ht}} \nonumber\\
&+\frac{\mathcal{B}_k^*}{\sqrt{2k}}\left. e^{-ik/H}\left(\frac{iH}{k} + e^{-Ht}\right)e^{ik/H\,e^{-Ht}} \right].
\end{align}
This can be recast as
\begin{align}
\mathscr{A}_\textbf{k} =  \frac{\mathcal{M}\mathcal{A}_k^*}{\sqrt{2k}}e^{ik/H}&\left(-\frac{iH}{k}\right) \lim_{\mu\rightarrow1}\left(1- \partial_\mu \right) I_\mu(\nu) \nonumber\\
&+ \frac{\mathcal{M}\mathcal{B}_k^*}{\sqrt{2k}}e^{-ik/H}\left(\frac{iH}{k}\right) \lim_{\mu\rightarrow1}\left(1- \partial_\mu \right) I_\mu^*(-\nu) 
\end{align}
where
\begin{align}
I_{\mu}(\nu)&= \int^T _{-T}dt\,e^{i\nu t}e^{-ik\mu/H\,e^{-Ht}} \nonumber\\
&= \left(\int ^{\infty} _{-\infty}dt\, - \int ^{-T} _{-\infty}dt\, -\int ^{\infty} _{T}dt\,\right)e^{i\nu t}e^{-ik\mu/H\,e^{-Ht}}.
\end{align}
The above integral can be evaluated to give,
\begin{equation}
I_{\mu}=\frac{1}{H}\left(\frac{k}{H}\right)^{i\nu/H} e^{-\pi \nu/2H} e^{\frac{i\nu}{H} \mathrm{ln} \mu} \left[\Gamma\left(-\frac{i\nu}{H},i\frac{k\mu}{H}e^{-HT}\right)-\Gamma\left(-\frac{i\nu}{H},i\frac{k\mu}{H}e^{HT}\right)\right]
\end{equation}
where $\Gamma(a,b)$ is an incomplete gamma function. With this the amplitude becomes,
\begin{equation}
\mathscr{A}_\textbf{k}= \frac{\mathcal{M}e^{ik/H}}{\sqrt{2k}} \left(\frac{-iH}{k}\right)\mathcal{A}_k^*\left(1-\frac{i\nu}{H}\right) I_1(\nu) + \frac{\mathcal{M}e^{-ik/H}}{\sqrt{2k}} \left(\frac{iH}{k}\right)\mathcal{B}_k^*\left(1-\frac{i\nu}{H}\right) I_1^*(-\nu)
\end{equation}
where we have ignored terms coming from differentiating the gamma functions since those are purely oscillatory and can be made to vanish in the large $\nu$ limit by using the standard $i\epsilon$ prescription.
The probability $P_\mathbf{k}$ for the transition is now given by,
\begin{align}
P_\mathbf{k}=\vert \mathscr{A}_\textbf{k}\vert ^2 = &\frac{\mathcal{M}^2H^2}{2k^3} \left(1+\frac{\nu^2}{H^2}\right)\left(\vert \mathcal{A}_k\vert^2\vert I_1(\nu)\vert ^2 + \vert \mathcal{B}_k\vert^2 \vert I_1(-\nu)\vert ^2\right.\nonumber\\
 &\left.+ 2\vert \mathcal{A}_k\vert\vert \mathcal{B}_k\vert\vert I_1(\nu)\vert \vert I_1^*(-\nu)\vert\,\cos\theta \right)
\end{align}
To avoid the transients arising due to finite $T$, we will take the limit of $HT\gg 1$. In this case, an elementary computation gives:   
\begin{equation}
 I_1 \approx \frac{e^{-\pi \nu/2H}}{H}\left(\frac{k}{H}\right)^{i\nu/H} \Gamma \left(-\frac{i\nu}{H}\right) - \frac{i}{k}e^{-HT}e^{-i\frac{k}{H} e^{HT}}e^{-i\nu T}.
\end{equation}
so that 
\begin{align}
 \vert I_1(\nu)\vert^2 &= \frac{\beta N}{\nu}  - \frac{2e^{-HT}}{k}\sqrt{\frac{\beta N}{\nu}} \cos\theta' + \mathcal{O}(e^{-2HT})\nonumber\\
 \vert I_1^*(-\nu)\vert^2 &= \frac{\beta e^{\beta\nu} N}{\nu}  + \frac{2e^{-HT}e^{\beta\nu/2}}{k}\sqrt{\frac{\beta N}{\nu}}\cos\theta'' + \mathcal{O}(e^{-2HT})
\end{align}
where $N = (e^{\beta\nu} -1)^{-1}$ and $\beta = 2\pi/H$. Therefore, when  $HT \gg 1$, we get the transition probability to be,
\begin{equation}
 P_\mathbf{k} = \frac{\mathcal{M}^2H^2}{2k^3} \left(1+\frac{\nu^2}{H^2}\right)\frac{\beta}{\nu}
\left[\left(\vert\mathcal{A}_k\vert^2 + \vert \mathcal{B}_k\vert^2e^{\beta\nu}\right) N + 2\vert \mathcal{A}_k\vert\vert \mathcal{B}_k\vert\sqrt{N(N+1)}\cos\theta \right]
\label{res2}
\end{equation}
A comparison  with \eq{res1} shows that the detector response is triggered by essentially $|f(-\nu)|^2$ which should be obvious from the fact that the amplitude in \eq{amp} picks out the negative frequency component of the field when the time integration is extended over the range $(-\infty,\infty)$.
This result shows that our mixing coefficients have a direct connection with the operational definition of particle content, as determined by the detector response. 

The above result is general and is valid for arbitrary $\mathcal{A}_k,\mathcal{B}_k$. By taking specific values we can determine the detector response in Bunch-Davies and co-moving vacuum. In the Bunch-Davies case, we have $\mathcal{A}_k=1$ and $\mathcal{B}_k=0$ giving
\begin{equation}
 P_\mathbf{k} = \frac{\mathcal{M}^2H^2}{2k^3} \left(1+\frac{\nu^2}{H^2}\right)\frac{\beta}{\nu} N 
 \label{response}
\end{equation}
This result shows that the detector response does pick up the extra factor $(1+\nu^2/H^2)$ just as the mixing coefficients do. (The same factor has been noticed earlier in ref. \cite{sspaper}). The corresponding result for  co-moving vacuum can be obtained by substituting \eq{ABcm} into \eq{res2} but the result has no special features worth mentioning.
The above results arise because, by definition, the geodesic detector measures the co-moving time $t$.


\section{Relation to the results in static coordinate system}
\label{sec:static}

In Sec.~\ref{sec:thermality}, we briefly described how thermal nature of the de Sitter horizon arises in the static coordinate system and in the last few sections we studied the field theory in Friedmann coordinate system. Since both coordinate systems coexist in part of the de Sitter manifold, one can make an explicit comparison of the quantum states defined in these two coordinate systems. (This is similar to comparing the states in inertial coordinate system and Rindler coordinate system in flat spacetime.) For this, we need to compute the relevant Bogoulibov coefficients on a spacelike hypersurface between the relevant mode functions by using the Klein-Gordon inner product. As we shall see, this is fairly straightforward in (1+1) but somewhat complicated in (1+3).


\subsection{Comparison in $\mathbf{dS_2}$}

We begin by noting that the metric,
\begin{equation}
ds^2 = dt^2 - e^{2Ht}dx^2
\end{equation}
in $(t,x)$ coordinates can be written in the static coordinates $(\tilde{t},\tilde{x})$ as
\begin{align}
ds^2 &= \left(1- H^2 \tilde{x}^2\right)d\tilde{t}\,^2 - \left(1- H^2 \tilde{x}^2\right)^{-1}d\tilde{x}^2\nonumber\\
&= \left(1- H^2 \tilde{x}^2(x_*)\right)(d\tilde{t}\,^2 - d\tilde{x}_*^2)
\label{cds}
\end{align}
where
\begin{equation}
\tilde{x} =  e^{Ht} x\,;\hspace{10pt}\tilde{t} = t - 1/2H\ln\left(1 - H^2\tilde{x}^2\right)
\end{equation}
and 
\begin{equation}
\tilde{x}_* = \int \frac{\dd \tilde{x}}{\left(1- H^2 \tilde{x}^2\right)}.
\end{equation}
 is the tortoise coordinate. Using these transformations we can express the field modes in the time-dependent $\mathrm{dS}_2$ coordinates in terms of  the static coordinates. We will focus on a  fixed $({k}>0)$ mode so that the mode function
\begin{equation}
f_k (t,x) = \frac{1}{\sqrt{2k}} e^{-ik/H}e^{ikx}e^{ik/H\, e^{-Ht}}
\end{equation} 
 becomes
\begin{equation}
f_k(\tilde{t},\tilde{x}) =  \frac{1}{\sqrt{2k}}\exp\left(\frac{ike^{-H\tilde{t}}}{\left(1- H^2 \tilde{x}^2\right)^{1/2}} + \frac{ike^{-H\tilde{t}}\tilde{x}}{H\left(1- H^2 \tilde{x}^2\right)^{1/2}}\right) = \frac{1}{\sqrt{2k}}e^{i(k/H)\,e^{-Hu}}
\end{equation}
in static coordinates
where $u\equiv \tilde{t} - \tilde{x}_*$. In the static $\mathrm{de Sitter}$ patch, conformal flatness of the metric in \eq{cds} 
allows us to write down the solution to the field equation as $\exp(\pm i\omega u)$.  This allows the expansion 
\begin{equation}
\Phi^R_\omega = \frac{1}{\sqrt{2\omega}}\left(\hat{b}_\omega e^{-i\omega u} + \hat{b}^\dagger_\omega e^{i\omega u}\right) 
\end{equation}
etc. which is valid on the complete manifold. We now need to determine the Bogoliubov coefficients that relate the above two sets of field modes. These are given by the standard Klein-Gordon inner product. 
\begin{equation}
\beta_{\omega k} = -i \int_{\tilde{t}}\dd \tilde{x}_*\left(\Phi^R_\omega\partial_{\tilde{t}}\, f_k - f_k\,\partial_{\tilde{t}}\, \Phi^R_\omega \right)
\end{equation}
where the integral is over any spacelike hypersurface. Choosing $\tilde{t} = 0$ surface, the above integral over $\tilde{x}_*$ can be recast as:
\begin{equation}
\beta_{\omega k}  = \frac{- i}{\sqrt{2\omega}}\int_{-\infty}^{\infty}\dd u\left(e^{-i\omega u}\partial_u f_k - f_k\,\partial_u e^{-i\omega u} \right)
\end{equation}
Integrating the first term by parts gives,
\begin{align}
\label{bogobeta}
\beta_{\omega k}  &= \sqrt{2\omega} \int_{-\infty}^{\infty}\dd u\, e^{-i\omega u} f_k (u) +  f_k\,e^{-i\omega u}|_{-\infty}^{\infty} \nonumber\\
&=\sqrt{2\omega} \int_{-\infty}^{\infty}\dd u\, e^{-i\omega u} f_k (u) 
\end{align}
since the second term vanishes. Thus
\begin{align}
\beta_{\omega k}  &= \sqrt{\frac{\omega}{k}}\int_{-\infty}^{\infty}\dd u e^{-i\omega u} e^{ik/H\,e^{-Hu}} \nonumber\\
          &=  \sqrt{\frac{\omega}{k}}\left(\frac{1}{H}\right)\left(\frac{k}{H}\right)^{i\omega/H}\Gamma\left(-\frac{i\omega}{H}\right) e^{-\pi\omega/2H}.
\end{align}
We find that
 modulus $|\beta_{\omega k}|^2$ is again Planckian at temperature, $ H/2\pi$:
\begin{equation}
|\beta_{\omega k}|^2 = \frac{\beta}{k (e^{\beta\omega} - 1)};\hspace{5pt}\beta = \frac{2\pi}{H}.
\end{equation}
This shows that the Bunch-Davies vacuum (which is the same as the co-moving vacuum in (1+1) dimension) has a thermal character in the static patch bounded by the horizon. 

\subsection{Comparison in $\mathbf{dS_4}$}

The transformation from the Friedmann coordinates   to static coordinates goes through 
 in $\mathrm{dS_4}$ exactly in the same way as $\mathrm{dS_2}$. The metric
\begin{equation}
ds^2 = dt^2 - e^{2Ht}\left(dr^2 + r^2d\Omega^2\right)
\end{equation}
in $(t,r,\Omega)$ system can be written in the static coordinates $(\tilde{t},\tilde{r},\Omega)$ as
\begin{align}
ds^2 &= \left(1- H^2 \tilde{r}^2\right)d\tilde{t}\,^2 - \left(1- H^2 \tilde{r}^2\right)^{-1}d\tilde{r}^2 - \tilde{r}^2d\Omega^2\nonumber\\
&= \left(1- H^2 \tilde{r}^2(r_*)\right)(d\tilde{t}\,^2 - dr_*^2) - \tilde{r}^2(r_*)d\Omega^2
\end{align}
with the same transformations as before
\begin{equation}
\label{trans}
\tilde{r} =  e^{Ht} r\,;\hspace{10pt}\tilde{t} = t - 1/2H\ln\left(1 - H^2\tilde{r}^2\right)
\end{equation}
and defining the tortoise coordinate
\begin{equation}
r_* = \int \frac{\dd \tilde{r}}{\left(1- H^2 \tilde{r}^2\right)}.
\end{equation}
In the static coordinates, the field equation reads
\begin{equation}
\left[\frac{\partial^2}{\partial{\tilde{t}}^2} - \frac{f(\tilde{r})}{{\tilde{r}}^2}\frac{\partial}{\partial{\tilde{r}}}\left(\tilde{r}^2f(\tilde{r})\frac{\partial}{\partial{\tilde{r}}}\right) -  \frac{f(\tilde{r})\hat{L}^2}{r^2}\right]\Phi(\tilde{t},\tilde{r},\Omega) = 0
\end{equation}
where $f(\tilde{r}) = (1 - H^2\tilde{r}^2)$ and $\hat{L}$ is the standard angular Laplacian operator. Taking\\ 
$\Phi = \phi_l(\tilde{r})Y_{lm}(\Omega)e^{-i\omega\tilde{t}}/\tilde{r}$, 
we find that $\phi_l(\tilde{r})$ satisfies the equation:
\begin{equation}
-\omega^2\phi_l - \frac{f}{\tilde{r}}\frac{d}{d\tilde{r}}\left(\tilde{r}^2f\frac{d}{d\tilde{r}}\left(\frac{\phi_l}{\tilde{r}}\right)\right) - \frac{l(l+1)f}{\tilde{r}^2}\phi_l = 0.
\end{equation} 
Since $f(\tilde{r})$ vanishes at the horizon $\tilde{r} = 1/H$, only the \emph{s}-mode makes a dominant contribution  near the horizon and hence we will focus on $l=0$ mode.
For this mode, the wave equation becomes
\begin{equation}
\frac{d^2\phi}{dr_*^2} + \left(\omega^2 - \frac{ff'}{\tilde{r}}\right)\phi = 0
\end{equation}
where the prime denotes derivative with respect to $\tilde{r}$. Clearly, in the near horizon limit ($f\to 0$), the solutions behave as $\exp(\pm i\omega r_*)$. Thus near the past horizon, $\tilde{r}\rightarrow1/H$ and $\tilde{t}\rightarrow -\infty$,  the modes in the static coordinate system behave as $\exp(\pm i\omega v)$. 

On the other hand, the modes describing the Bunch-Davies vacuum can be expressed in spherical coordinates by the standard plane wave expansion
\begin{equation}
\Phi_k^{BD} = \frac{e^{-ik/H}}{\sqrt{2k}}\sum_{l=0}^{\infty} i^l(2l+1)j_l(kr)P_l(\cos\theta)e^{ik/H e^{-Ht}}\left(\frac{iH}{k} + e^{-Ht}\right)
\end{equation}
Using the transformations in \eq{trans} we can express this in $(\tilde{t},\tilde{r})$ coordinates. Concentrating on the \emph{s}-wave contribution we obtain 
\begin{equation}
\Phi_k^{(BD)} = \frac{e^{-ik/H}}{\sqrt{2k}}\left(e^{ik/H e^{-Hu}} - e^{ik/H e^{-Hv}}\right)\left(\frac{iH}{k}e^{H\tilde{t}}\left(1- H^2\tilde{r}^2\right) + 1\right). 
\end{equation}
which, near the past horizon, $\tilde{r}\rightarrow1/H$ and $\tilde{t}\rightarrow -\infty$, behaves as
\begin{equation}
\Phi_k^{(BD)} \rightarrow \frac{1}{\sqrt{2k}}e^{i(k/H)e^{-Hv}}.
\end{equation}
We now use the fact that the Klein-Gordon inner product between the field modes is independent of the surface over which it is evaluated. It is, therefore, convenient to evaluate the Bogoliubov coefficients on a space-like surface very close to the horizon.
Since the Bunch-Davies mode behaves as $ e^{i(k/H)e^{-Hv}}$ while the static modes behave as $\exp(\pm i\omega v)$, it is obvious that  the Bogoliubov coefficients defined in \eq{bogobeta} will give
\begin{equation}
\beta_{\omega k} = \sqrt{2\omega} \int_{-\infty}^{\infty}\dd v\, e^{-i\omega v} \Phi_k^{(BD)} (v) 
\end{equation}
which has a thermal character:
\begin{equation}
|\beta_{\omega k}|^2 = \frac{\beta}{k (e^{\beta\omega} - 1)};\hspace{5pt}\beta = \frac{2\pi}{H}.
\end{equation}
We again see that the Bunch-Davies vacuum has a thermal property when viewed in the static patch in (1+3) dimensions as well. In this sense, (1+1) and (1+3) dimensions behave identically. It is also straightforward to show that a detector at rest in the static coordinates will perceive a thermal radiation in the Bunch-Davies vacuum state. On the other hand, we saw earlier that a \textit{freely-falling} detector will also see the modified thermal spectrum [see \eq{response}] in the same vacuum state. It should be noted that this is somewhat contrary to the results in black hole spacetime.

Finally, we quote the result for the co-moving vacuum transformed to static coordinates. The analysis is again straightforward when we use the fact that  the co-moving modes can be expressed in terms of the Bunch-Davies modes by the relation
\begin{equation}
\psi^{(CM)}_k(t) = \mathcal{A}_k \psi^{(BD)} (t) + \mathcal{B}_k {\psi^{(BD)}}^* (t) 
\end{equation}
Therefore, 
\begin{equation}
\Phi_k^{(CM)} \rightarrow \frac{1}{\sqrt{2k}}\left(\mathcal{A}_k e^{ik/He^{-Hv}} + \mathcal{B}_k e^{-ik/He^{-Hv}}\right).
\end{equation}
on the past horizon. It follows that the spectrum is now given by 
\begin{align}
k|\beta_{\omega k}|^2 =  \left(|\mathcal{A}_k|^2 + |\mathcal{B}_k|^2 e^{\beta\nu}\right) \beta N + |\mathcal{A}_k||\mathcal{B}_k|\beta\sqrt{N(N+1)}\cos \theta
\end{align}
where
\begin{equation*}
N = \frac{1}{e^{\beta \nu} -1}
\end{equation*}
We once again see that the co-moving vacuum introduces an interference term in the form of $\sqrt{N(N+1)}$ even when we compare the modes between Friedmann description and static description suggesting that there must be some physical explanation for the origin of this factor. We hope to address this question in a future publication.


\section{Quantum fields in quasi-de Sitter spacetime}
\label{sec:qds}

So far, we have been concentrating on the features which are special to de Sitter spacetime.  However, in the evolution of the real universe, it is impossible to obtain a pure de Sitter evolution due to the presence of external matter. Both, during the inflationary phase as well as during the late time acceleration phase, we only have a quasi-de Sitter phase rather than a pure de Sitter universe. In this section we will extend the formalism described earlier to a quasi-de Sitter spacetime by determining an approximate solution to the wave equation. This approach is quite general and can take into account any first order deviation from the pure de Sitter universe. After developing the formalism we will apply it to a specific example to illustrate its utility.


\subsection{The perturbative framework}
\label{subsec:perturbativeframework}

Consider a Friedmann spacetime with the scale factor given by:
\begin{equation}
a(t)=e^{(Ht+\epsilon \lambda(t))}\approx e^{Ht}\left(1+\epsilon \lambda(t)\right) =  a_0 + \epsilon \lambda a_0
\end{equation}
which can be treated as quasi-de Sitter if the condition $\ddot{\lambda}\ll\dot{\lambda}H$ is satisfied. In the above expansion, we have retained the perturbation to first order as indicated by the bookkeeping parameter $\epsilon$ (which will be set to unity at the end of the computation).  Correspondingly, the mode functions, which are the solutions to the wave equation in the perturbed metric, will differ from those in the de Sitter spacetime by a small amount:
\begin{equation}
\psi(t) = \psi_0(t) + \epsilon\,\delta \psi(t)
\end{equation}
where $\psi_0(t)$ is the unperturbed mode function 
and we have omitted the subscript $k$ for notational simplicity. Substituting the above expressions for $a(t)$ and $\psi(t)$ into the time dependent part of the wave equation written in the form: 
\begin{equation}
\frac{d^2 \psi}{dt^2} + 3\left(\frac{\dot{a}}{a}\right)\frac{d\psi}{dt}+\frac{k^2}{a^2}\psi = 0.
\end{equation}
we get,
\begin{equation}
\frac{d^2(\delta \psi)}{dt^2}+3H\frac{d(\delta \psi)}{dt}+3\dot{\lambda}\frac{d\psi_0}{dt}-2\frac{k^2}{a_0^2}\lambda\psi_0 + \frac{k^2}{a_0^2}\delta \psi = 0.
\end{equation}
This equation can be solved by writing
$\delta \psi$ = $\psi_0\, s$.
The function $s(t)$ then satisfies the equation,
\begin{equation}
\frac{d^2 s}{dt^2} +\left(2 \frac{\dot{\psi}_0}{\psi_0}+3H\right)\frac{ds}{dt}= \mu (t)
\label{soft}
\end{equation}
where
\begin{equation}
\mu (t) \equiv 2\frac{k^2}{a_0^2}\lambda - 3\dot{\lambda}\frac{\dot{\psi}_0}{\psi_0}
\end{equation}
acts like a source term. \eq{soft} is first order in $ds/dt$ and hence can be immediately integrated. (This result holds for  a generic class of second-order homogeneous linear differential equation;  see \app{theorem} for details). The solution for $s(t)$ is
\begin{equation}
\label{s:gensol}
s (t) = C\int^t \mathrm{d}t' \psi_0^{-2} e^{-3Ht'}  + \int^{t} \mathrm{d}t' \, \psi_0^{-2}(t')\,e^{-3Ht'}\int^{t'} \mathrm{d}t'' \psi_0^2(t'') e^{3Ht''}\mu (t'')
\end{equation}
where $C$ is a constant of integration. Thus, given a model for $\lambda (t)$ and appropriate boundary conditions we can, in principle, solve for the perturbation $\delta \psi$ by this method.


\subsection{An example: late-time accelerated phase of the universe}
\label{subsec:latetime}

As an illustration of the above method, let us consider the late time accelerated phase of the universe containing dust-like matter and a cosmological constant. The expansion factor of such a universe is given by 
\begin{equation}
a(t) = 2^{2/3} \left(\sinh \frac{3}{2} Ht\right)^{2/3}
\end{equation}
In the spirit of the above discussion, we will treat this as a perturbation to an exact de Sitter universe and write
\begin{equation}
a(t)\approx e^{Ht}\left(1 - \frac{2}{3}e^{-3Ht}\right)
\end{equation}
where $\lambda (t) = -(2/3)\exp(-3Ht)$ which vanishes as $t$ goes to infinity. 
This behaviour suggests that we use the boundary conditions
  $s(\infty) = 0$ and $\dot{s}(\infty) = 0$ 
in our general solution given by \eq{s:gensol}. In the pure de Sitter case, the mode functions can be taken to be 
\begin{equation}
\psi_0(t)=\frac{1}{\sqrt{2k}}\exp\left[-\frac{ik}{H}\left(1- e^{-Ht}\right)\right]\left(\frac{iH}{k} + e^{-Ht}\right)
\end{equation}
which amounts to taking $\mathcal{A}_\textbf{k}=1$ and $\mathcal{B}_\textbf{k}=0$ in the \eq{psi_0} i.e., we have chosen to work with Bunch-Davies state.
Calculating the integrals in  \eq{s:gensol} is straightforward (see \app{s:latetimesol} for details) and we obtain,
\begin{equation}
s(t) = \frac{7k^2}{15H^2}e^{-5Ht}
\end{equation} 
 Therefore, the first-order change in the mode function is given by:
\begin{equation}
\delta\psi (t) = \psi_0(t)s(t) = \frac{7ik\,e^{-ik/H}}{15H\sqrt{2k}}\left(1 - \frac{i k}{H}e^{-Ht}\right) e^{-5Ht}e^{\frac{ik}{H}e^{-Ht}}.
\end{equation} 
We can now compute the Fourier transform of this expression to determine the first order correction in Fourier space: 
\begin{align}
\delta f (\nu) =& \int_{-\infty}^{\infty} \mathrm{d}t\,\delta \psi(t)e^{i\nu t}\nonumber\\
=& \frac{7ik}{15 H\sqrt{2k}} e^{-ik/H}  \int_{-\infty}^{\infty} \mathrm{d}t\, \left(1 - \frac{i k}{H}e^{-Ht}\right) e^{-5Ht}e^{\frac{ik}{H}e^{-Ht}} e^{i\nu t}\nonumber\\
=& \frac{7ik}{15 H\sqrt{2k}} e^{-ik/H} \lim_{\mu\rightarrow 1}(1 - \partial_\mu) \int_{-\infty}^{\infty} \mathrm{d}t\, e^{-5Ht + i\nu t}e^{\frac{ik\mu}{H}e^{-Ht}}\nonumber\\
=& \frac{-7H^3\,e^{-ik/H}}{15k^4\sqrt{2k}}\left(\frac{k}{H}\right)^{i\nu/H}e^{\pi\nu/2H}\left(6-\frac{i\nu}{H}\right)\Gamma\left(5-\frac{i\nu}{H}\right)
\end{align}
The resulting power spectrum, to the lowest order, is given by:
\begin{equation}
\nu |F(\nu)|^2 \approx \nu |f(\nu)|^2 +  \epsilon \,\nu \mathrm{Re}[2f^*(\nu)\delta f(\nu)]
\end{equation}
where we have reintroduced $\epsilon$ for bookkeeping and 
\begin{equation}
\nu\mathrm{Re}[2f^*(\nu)\delta f(\nu)] = \frac{7H^5}{15k^6}\frac{\beta e^{\beta\nu}}{e^{\beta\nu} - 1}\left(144\frac{\nu}{H} + 64 \frac{\nu^3}{H^3} - 79\frac{\nu^5}{H^5} + \frac{\nu^7}{H^7}\right)
\end{equation}
is the \textit{correction} to the power spectrum in the case of quasi-de Sitter phase arising  from the matter contribution in the  late-time acceleration.


\section{Conclusions}\label{sec:conclusions}

The periodicity in the Euclidean time allows us to attribute a temperature $H/2\pi$ using the static coordinates on the de Sitter manifold. In this sense, de Sitter spacetime behaves just like other static spacetimes with a horizon. However, in such an analysis, one has to define a vacuum state on a $T=0$ hypersurface in the Kruskal-like coordinate system  which is not static. In this particular case, thermal nature of the de Sitter horizon arises because the vacuum state in Kruskal-type coordinate system leads to a thermal density matrix for the observers bounded by the de Sitter horizons.

The de Sitter spacetime is unique in the sense that it also allows introducing Friedmann coordinates in which the metric is homogeneous. This, in turn, reduces the field theoretic problem to that of a quantum oscillator with a time dependent frequency. In such a, time dependent, background there is no unique definition for the vacuum state and the best one could do is to introduce well motivated vacuum states and study their physical properties. Quite generically, such states can be introduced by giving a suitable boundary condition for the mode functions at some time $t=t_0$. The question arises as to whether one can understand the thermality of de Sitter universe working \emph{entirely within} the Friedmann coordinates i.e.,\textit{ without} comparing the results between Friedmann and static coordinates. 
(We have not seen such a derivation in the literature for a massless scalar field.)
We investigated several aspects of this question both in (1+1) dimension and in (1+3) dimensions in this paper.

Two natural vacuum states one can introduce are the Bunch-Davies and co-moving vacuum states in this spacetime. In (1+1) dimension, the Bunch-Davies vacuum state corresponds to choosing the modes to be positive frequency with respect to the conformal time $\eta$ in the asymptotic past while the co-moving vacuum state corresponds to imposing the positive frequency condition at some arbitrary instant of time $t=t_0$.  It turns out that both these states are identical in (1+1) dimension. To study the time evolution of this state, we expand the mode function in terms of positive and negative frequency modes defined with respect to the co-moving time. The mixing of positive and negative frequency modes then reveals a thermal  character with temperature $H/2\pi$.   This, of course, does not happen during the time evolution in the conformal time; the positive frequency mode remains a positive frequency mode at all times.

The situation in (1+3) dimensions is quite different. To begin with, co-moving vacuum state and the Bunch-Davies vacuum state do not coincide in (1+3) dimensions.  Further, the mixing coefficient between positive and negative frequency modes does not have a pure thermal character (and is modified by an extra frequency dependent factor) in the case of Bunch-Davies vacuum. The result for the case of co-moving vacuum is more complicated and involves an interference term containing $\sqrt{N(N+1)}$ factor which is reminiscent of the fluctuations in the occupation numbers of massless thermal radiation. 

The physical meaning of the mixing coefficients introduced to analyse the above phenomena can be understood by studying the response of particle detectors in the de Sitter spacetime.  We computed the rate of excitation of a geodesic detector evolving in co-moving time. This rate exactly matches with the particle content of the state as determined by the mixing coefficients in both Bunch-Davies vacuum and co-moving vacuum. 

We also compared the states defined using Friedmann coordinate system with those defined using the static coordinate system. This requires evaluating the necessary Bogoliubov coefficient between the mode functions defined in the static patch and Friedmann patch in the region of the manifold where they coexist. We found that the Bunch-Davies vacuum appears to be a thermal state for static observers bounded by the horizon, both in (1+1) and (1+3) dimensions. This is in contrast with the results obtained within the Friedmann coordinate system where the results for (1+3) dimensions \textit{differ} from the results for (1+1). On the other hand, the co-moving vacuum in (1+3) dimension, defined in Friedmann coordinates, does not have a simple thermal interpretation in the static coordinates. 

In the last part of the paper, we studied the effects of small deviations from de Sitter evolution and the resulting corrections to the mode functions. This formalism is sufficiently general to handle any functional form of the deviation in the lowest order of perturbation theory. As an illustration of this formalism, we studied  the deviations in the power spectrum arising due to the existence of pressure-free matter during the late time accelerated phase of the universe. This formalism might have applications for studying the spectral deviations in the case of inflationary universe as well.


\section*{Acknowledgements}

SS is supported by a fellowship from the Council of Scientific and Industrial Research (CSIR), India. CG would like to thank IUCAA for hosting her in the VSP program. TP's research is partially supported by the J.C.Bose Research Grant of DST, India.

\appendix{}

\section{Calculation of the Fourier transform in \eq{eq:ft}}
\label{ft}
To evaluate
\begin{equation*}
I =\int ^{\infty} _{-\infty}dt\,e^{i\nu t}\,e^{\frac{ik \mu}{H}e^{-Ht}},
\end{equation*}
we define $u=e^{-Ht}$ and $ b = k\mu/H$. This gives,
\begin{align}
I &=\frac{1}{H}\int _0 ^{\infty} du\, u^{-1-\frac{i\nu}{H}} e^{ibu}\nonumber\\
 &=\frac{1}{H} \exp \left[\frac{i\nu}{H} \mathrm{ln} \left \vert \frac{k\mu}{H} \right \vert +\frac{\pi \nu}{2H} \mathrm{sign} \left(\frac{k\mu}{H}\right)\right]\Gamma \left(-\frac{i\nu}{H}\right)\nonumber\\
&=\frac{1}{H} \left(\frac{k}{H}\right)^{i\nu/H}e^{\pi \nu/2H}\Gamma \left(-\frac{i\nu}{H}\right) e^{\frac{i\nu}{H} \mathrm{ln} \mu}.
\end{align}

\section {A result in perturbation theory}
\label{theorem}
Consider a generic second-order homogeneous linear differential equation
\begin{equation*}
a(t) \ddot{x}(t) + b(t) \dot{x}(t) + c(t) x(t) = 0
\end{equation*} 
Let $x_0$ be the solution of above equation for some functions $a_0(t)$, $b_0(t)$ and $c_0(t)$. We are now interested in the corresponding solution of the equation when the parameter functions $a$, $b$ and $c$ are perturbed about their original forms. Then, to the first order in perturbation, we have
\begin{equation}
a_0 \delta\ddot{x} + \ddot{x}_0\delta a + b_0 \delta\dot{x} + \dot{x}_0\delta b + c_0\delta x + x_0\delta c = 0.
\end{equation}
Scaling the perturbation $\delta x$ with the unperturbed solution as, $\delta x \equiv x_0 s$, gives for $s(t)$ the equation 
\begin{equation}
\ddot{s} + A(t) \dot{s} = B(t)
\end{equation}
with
\begin{align}
A(t) &= 2\frac{\dot{x}_0}{x_0} + \frac{b_0}{a_0}\nonumber\\
B(t) &= -\frac{\ddot{x}_0}{x_0}\frac{\delta a}{a_0} - \frac{\dot{x}_0}{x_0}\frac{\delta b}{a_0} - \frac{\delta c}{a_0}
\end{align}
which is a  first order differential equation in $\dot{s}$ and can be solved immediately. 

\section {Solution of \eq{s:gensol} for late-time accelerated phase}
\label{s:latetimesol}

The basic ingredients that go in are
\begin{equation}
\psi_0= \frac{1}{\sqrt{2k}}\exp\left[-\frac{ik}{H}\left(1- e^{-Ht}\right)\right]\left(\frac{iH}{k} + e^{-Ht}\right),\nonumber
\end{equation}
which gives
\begin{equation}
\frac{\dot{\psi}_0}{\psi_0} = \frac{-\,ik^2e^{-2Ht}}{H\left(i+\frac{k}{H}e^{-Ht}\right)}\nonumber
\end{equation}
and
\begin{align}
\mu(t) &= 2k^2 e^{-2Ht} \lambda - 3\dot{\lambda}\frac{\dot{\psi}_0}{\psi_0} = -\frac{4}{3}k^2 e^{-5Ht} + \frac{6i k^2e^{-5Ht}}{\left(i + k/H\, e^{-Ht}\right)}\nonumber
\end{align}
so that under the conditions $s(\infty)=\dot{s}(\infty)=0$, we can set $C$, the constant of integration in the homogeneous part to be zero and  obtain:
\begin{align}
s(t) = & \left(\frac{H^2}{6k}\right)e^{-\frac{2ik}{H}}\int_{t}^{\infty}\mathrm{d}t'\, \frac{e^{-3Ht'}}{\psi_0^2(t')}\int_{t'}^{\infty}\mathrm{d}t''\,e^{\frac{2ik}{H}\,e^{-Ht''}}e^{-2Ht''}\left[-14 - 4\frac{k^2}{H^2}e^{-2Ht''} + 10 \frac{ik}{H}e^{-Ht''}\right]\nonumber\\
 = &\left(\frac{H}{12}\right)\int_{t}^{\infty}\mathrm{d}t'\,e^{-\frac{2ik}{H}\,e^{-Ht'}}e^{-3Ht'}\left[-14\gamma\left(2,-2i\frac{k}{H}e^{-Ht'}\right)+\gamma\left(4,-2i\frac{k}{H}e^{-Ht'}\right)- \right.\nonumber\\ 
&\left.5\gamma\left(3,-2i\frac{k}{H}e^{-Ht'}\right)\right]\nonumber
\end{align}
Noting that,
\begin{equation*}
-14\gamma\left(2,x\right) + \gamma\left(4,x\right) - 5\gamma\left(3,x\right) = -18 + e^x(18 - 18x + 2 x^2 + x^3)
\end{equation*}
we can evaluate $s(t)$ in the late-time approximation to give
\begin{equation}
s(t) = \frac{7k^2}{15H^2}e^{-5Ht}
\end{equation}
as the leading order correction term.

\end{document}